\documentstyle[11pt,epsfig]{article}
\title{\bf Classical and quantum spinor cosmology with signature change}
\author{B. Vakili\thanks{email:
b-vakili@cc.sbu.ac.ir}, S. Jalalzadeh\thanks{email:
s-jalalzadeh@cc.sbu.ac.ir}
  and H. R. Sepangi\thanks{email:
hr-sepangi@cc.sbu.ac.ir}
\\ {\small Department of Physics, Shahid Beheshti University, Evin,
Tehran 19839, Iran}\\}
\begin{document}
\maketitle %\baselineskip 24pt

\begin{abstract}
We study the classical and quantum cosmology of a universe in
which the matter source is a massive Dirac spinor field and
consider cases where such fields are either free or
self-interacting. We focus attention on the spatially flat
Robertson-Walker cosmology and classify the solutions of the
Einstein-Dirac system in the case of zero, negative and positive
cosmological constant $\Lambda$. For $\Lambda<0$, these solutions
exhibit signature transitions from a Euclidean to a Lorentzian
domain. In the case of massless spinor fields  it is found that
signature changing solutions do not exist when the field is free
while in the case of a self-interacting spinor field such
solutions may exist. The resulting quantum cosmology and the
corresponding Wheeler-DeWitt equation are also studied for both
free and self interacting spinor fields and closed form
expressions for the wavefunction of the universe are presented.
These solutions suggest a quantization rule for the energy.
\vspace{5mm}
\end{abstract}\pagebreak

\section{Introdution}
The study of cosmology has always been influenced by the choice of
the matter field used to construct the energy-momentum tensor in
Einstein field equations. The most widely used matter source has
traditionally been the perfect fluid. However, the ubiquitous {\it
scalar field} has also been playing an increasingly important role
in more recent cosmological models as the matter source. This of
course is not surprising since being a ``scalar field'' makes it
somewhat easy to work with. One may also conceivably imagine a
universe filled with massless or massive {\it spinor fields} as
the matter source. Such cosmological models  have seldom been
studied in the literature, and more often than not, in the form of
general formalisms. In general then, it would be fair to say that
cosmologies with spinor fields as the matter source are the least
studied scenarios.

A question of interest related to classical and quantum
cosmological models is that of signature transition which has been
attracting attention since the early 1980's. Traditionally, a
feature in general relativity is that one usually fixes the
signature of the space-time metric before trying to solve
Einstein's field equations. However, there is no {\it a priori}
reason for doing so and it is now well known that the field
equations do not demand this property, that is, if one relaxes
this condition one may find solutions to the field equations which
when parameterized suitably, can either have Euclidean or
Lorentzian signature.

The notion of signature transition mainly started to appear in the
works of Hartle and Hawking \cite{1} where they argued that in
quantum cosmology amplitudes for gravity should be expressed as
the sum of all compact Riemannian manifolds whose boundaries are
located at the signature changing hypersurface. In more recent
times, a number of authors have studied this problem when a scalar
field is coupled to Einstein field equations and shown that the
resulting solutions, when properly parameterized, exhibit
signature transition, see for example \cite{2}. In a similar vain,
a classical model is studied in \cite{3} in which a
self-interacting scalar field is coupled to Einstein's equations
with a Sinh-Gordon interaction potential. The field equations are
then solved exactly for the scale factor and scalar field, giving
rise to a spatially flat Robertson-Walker cosmology with signature
changing properties. The case of the non-flat universe is
addressed in \cite{4} with a discussion about the conditions under
which signature transition exists. Also, it is shown in \cite{4}
that in the case of a massless scalar field this phenomena does
not exist. At the quantum cosmology level the same problem is
investigated in \cite{5} with an analysis pertaining to the exact
solutions of the Wheeler-DeWitt equation. Signature transition has
also been studied in multi-dimensional classical and quantum
cosmology in \cite{6} where a $4+d$-dimensional space-time is
minimally coupled to a scalar field. Finally, it has been used as
the compactification mechanism in Kaluza-Klein cosmology
\cite{7,8} for a positive and negative cosmological constant
respectively.

As a natural extension of the works done in \cite{3} and \cite{4}
and as a specific example, we consider a spinor field as the
matter source interacting with gravity and itself in a
Robertson-Walker geometry. This is the subject of study in this
paper. In general, studying spinor fields coupled to gravity
results in Einstein-Dirac systems which are not easy to solve. The
cosmological solutions of Einstein-Dirac systems have been studied
in few cases by some authors, see \cite{9} for example and the
references therein. The fourth reference in \cite{9} is notable in
that the quantization of a spinor field coupled to gravity is
studied in a Robetrson-Walker background. Here we analyze a
classical model with spatially flat Robertson-Walker cosmology in
which a massive spinor field is coupled to gravity. We show that
in this model signature transition occurs only if the cosmological
constant is negative. We also show that a massless free spinor
field does not lead to signature transition while a massless
self-interacting spinor field does.

The quantum cosmology of this model, embodied in the solutions of
the corresponding Wheeler-DeWitt (WD) equation is studied for both
free and self interacting spinor fields. It turns out that the WD
equation posses exact solutions in terms of  Hypergeometric
functions. Also, the ``zero energy condition" imposes a
quantization condition on the solutions for  the spinor fields.
These wavefunctions  correspond to the classical solutions
undergoing signature transition from a Euclidean to a Lorentzian
domain and, as such, could be useful in understanding  the initial
condition of the universe.

The paper is organized as follows: in section 2 we write the Dirac
equation in spatially flat Robertson-Walker background. Section 3
deals with the energy-momentum tensor of the spinor field and the
Einstein-Dirac  field equations. In section 4 we classify the
exact solutions in the cases of a free and self-interacting spinor
field when the cosmological constant is zero, negative or positive
respectively and investigate signature transition in each case.
The effect of a massless spinor field on signature transition is
studied in section 5. Section 6 is devoted to the study of the
quantum cosmology of our model and finally, conclusions are drawn
in section 7.
\section{Dirac equation}
Since our aim is to solve the coupled Einstein-Dirac equations in
a cosmological background, let us concentrate on writing the Dirac
equation in a spatially flat Robertson-Walker space-time given by
\begin{equation} \label{A}
ds^2=-dt^2+R^2(t)(dx^2+dy^2+dz^2).
\end{equation}
As is well known the Dirac equation in  curved space-time can be
obtained from the Lagrangian ($c=\hbar=1$)
\begin{eqnarray} \label{B}
L=\frac{1}{2}\left[\bar{\psi}\gamma^{\mu}(\partial_{\mu}+\Gamma_{\mu})\psi-
\bar{\psi}(\overleftarrow{\partial_{\mu}}-\Gamma_{\mu})\gamma^\mu\psi
\right]-V(\bar{\psi},\psi),
\end{eqnarray}
where $V(\bar{\psi},\psi)$ is a potential describing the
interaction of the spinor field $\psi$ with itself, $\gamma^{\mu}$
are the Dirac matrices associated with the spacetime metric
satisfying $\left\{\gamma^{\mu},\gamma^{\nu}\right\}=2g^{\mu \nu}$
and $\Gamma_{\mu}$ are the spin connections. The Euler-Lagrange
equations for $\psi$ and $\bar{\psi}$ then yield
\begin{equation} \label{C}
\gamma^{\mu}(\partial_{\mu}+\Gamma_{\mu})\psi-\frac{\partial
V}{\partial \bar{ \psi}}=0,
\end{equation}
\begin{equation} \label{D}
\bar{\psi}(\overleftarrow{\partial_{\mu}}-\Gamma_{\mu})\gamma^{\mu}+\frac{\partial
V}{\partial \psi}=0.
\end{equation}
The $\gamma^{\mu}$ matrices are related to the flat Dirac
matrices, $\gamma^a$, through the tetrads $e^a_{\mu}$ as follows
\begin{equation} \label{E}
\gamma^{\mu}=e^{\mu}_{a}\gamma^a,\hspace{.5cm}
\gamma_{\mu}=e^{a}_{\mu}\gamma_a.
\end{equation}
For metric (\ref{A}) the tetrads can be easily obtained from their
definitions, that is $g_{\mu\nu}=e^a_{\mu}e^b_{\nu}\eta_{ab}$,
leading to
\begin{equation} \label{F}
e^a_{\mu}=\mbox{diag}(1,R,R,R),\hspace{.5cm}
e^{\mu}_a=\mbox{diag}(1,1/R,1/R,1/R).
\end{equation}
Also, the spin connections satisfy the relation
\begin{equation} \label{G}
\Gamma_{\mu}=\frac{1}{4}g_{\nu\lambda}\left(\partial_{\mu}e^{\lambda}_a+
\Gamma^{\lambda}_{\sigma\mu}e^{\sigma}_a\right)\gamma^{\nu}\gamma^a.
\end{equation}
Thus, for the line element (\ref{A}), use of (\ref{E}) and
(\ref{F}) yeilds
\begin{equation} \label{H}
\Gamma_0=0,\hspace{.5cm}
\Gamma_i=-\frac{\dot{R}}{2}\gamma^{0}\gamma^{i},
\end{equation}
where $\gamma^0$ and $\gamma^i$ are the Dirac matrices in the
Minkowski space and we have adapted the following representation
\begin{equation} \label{I}
\gamma^0=\left(\!\!\!\begin{array}{cc} -i & 0
\\ 0 & i
\end{array}\!\!\!\right),\hspace{.5cm}\gamma^i=\left(\!\!\!\begin{array}{cc} 0 &
\sigma^i
\\ \sigma^i & 0
\end{array}\!\!\!\right).
\end{equation}
The preliminary setup is now complete for writing the Dirac
equation. From equations (\ref{C}) and (\ref{D}) we have
\begin{equation} \label{J}
\left[\frac{\partial}{\partial
t}-\frac{1}{R}\gamma^{0}\left(\gamma^1\frac{\partial}{\partial
x}+\gamma^2\frac{\partial}{\partial
y}+\gamma^3\frac{\partial}{\partial
z}\right)+\frac{3}{2}\frac{\dot{R}}{R}\right]\psi+\gamma^{0}\frac{\partial
V}{\partial \bar{\psi}}=0,
\end{equation}
\begin{equation} \label{L}
\bar{\psi}\left[\frac{\overleftarrow{\partial}}{\partial
t}-\frac{1}{R}\left(\frac{\overleftarrow{\partial}}{\partial
x}\gamma^1+\frac{\overleftarrow{\partial}}{\partial
y}\gamma^2+\frac{\overleftarrow{\partial}}{\partial
z}\gamma^3\right)\gamma^0+\frac{3}{2}\frac{\dot{R}}{R}\right]-\frac{\partial
V}{\partial \psi}\gamma^0=0.
\end{equation}
Since the metric functions in (\ref{A}) are functions of $t$ only,
we may write $\psi({\bf r},t)=e^{i{\bf p}\cdot{\bf r}}\psi(t)$ in
(\ref{J}) where ${\bf p}=(p_1,p_2,p_3)$ are the separation
constants.

The energy-momentum tensor for the Dirac spinor field as a matter
source for gravity can be obtained from the definition
\begin{equation} \label{LL}
T_{\mu \nu}=2\frac{\partial  L}{\partial g^{\mu\nu}}-g_{\mu\nu}L,
\end{equation}
which results in
\begin{equation}\label{M}
T_{\mu\nu}=\frac{1}{2}\left[\bar{\psi}\gamma_{( \mu}\nabla_{\nu
)}\psi-\left(\nabla_{(\mu}\bar{\psi}\right)\gamma_{\nu )}\psi
\right]-g_{\mu\nu} L,
\end{equation}
where
$$\nabla_{\mu}\psi=(\partial_{\mu}+\Gamma_{\mu})\psi\hspace{5mm}
\mbox{and}\hspace{5mm}
\nabla_{\mu}\bar{\psi}=\bar{\psi}(\overleftarrow{\partial_{\mu}}-\Gamma_{\mu}).$$
Since the Einstein tensor associated with  metric (\ref{A}) is
diagonal, the energy-momentum tensor of its matter source should
also be diagonal. With the anzats $\psi({\bf r},t)=e^{i{\bf
p}\cdot{\bf r}}\psi(t)$ the off-diagonal components of
$T_{\mu\nu}$ become
$$T_{ij}=\frac{1}{2}(p_{i}\bar{\psi}\gamma_{j}\psi+p_{j}\bar{\psi}\gamma_{i}\psi)
\hspace{5mm}\mbox{and}\hspace{5mm}
T_{0j}=-p_{j}\bar{\psi}\gamma_{0}\psi,$$ and hence for them to
vanish we must have $p_i=0$, {\it i.e.} the spinor field $\psi$
depends on $t$ only. Therefore  equations (\ref{J}) and (\ref{L})
simplify to
\begin{equation} \label{N}
\frac{d
\psi}{dt}+\frac{3}{2}\frac{\dot{R}}{R}\psi+\gamma^{0}\frac{\partial
V}{\partial \bar{\psi}}=0,
\end{equation}
\begin{equation} \label{O}
\frac{d
\bar{\psi}}{dt}+\frac{3}{2}\frac{\dot{R}}{R}\bar{\psi}-\frac{\partial
V}{\partial \psi}\gamma^0=0.
\end{equation}
The components of the energy-momentum tensor can then be evaluated
from  equations (\ref{M}) and (\ref{N}) with the result
\begin{equation} \label{P}
T^0_0=V(\bar{\psi},\psi), \hspace{.5cm}
T^1_1=T^2_2=T^3_3=-\frac{1}{2}\left(\bar{\psi}\frac{\partial
V}{\partial \bar{\psi}}+\frac{\partial V}{\partial
\psi}\psi\right)+V(\bar{\psi},\psi).
\end{equation}
\section{Field equations}
In this section we consider the Einstein-Dirac equation written as
\begin{equation} \label{Q}
G_{\mu\nu}=R_{\mu\nu}-\frac{1}{2}{\cal R}g_{\mu\nu}+\Lambda
g_{\mu\nu}=\kappa T_{\mu\nu}(\bar{\psi},\psi),
\end{equation}
where $R_{\mu\nu}$, ${\cal R}$ and $\Lambda$ are Ricci tensor,
scalar curvature and cosmological constant respectively,
constructed from torsion-free connections compatible with the
metric. The spinor field $\psi$ which interacts with itself
through the potential $V(\bar{\psi},\psi)$ is a solution of the
Dirac equation (\ref{N}) and is  coupled to gravity with the
energy-momentum tensor given by equation (\ref{M}). These coupled
Einstein-Dirac equations must now be solved in a domain that would
lead to a spatially flat Robertson-Walker cosmology given by
equation (\ref{A}) with Lorentzian signature $(-,+,+,+)$. However
we may parameterize the metric in such a way as to allow solutions
with continuous transition to a Euclidean domain. To this end, we
parameterize the metric, as in \cite{3} by  adapting the chart
$\{\beta ,x,y,z\}$ where the hypersurface of signature change
would be characterized by $\beta=0$. The metric can then be
parameterized in terms of the scale function $R(\beta)$ and the
lapse function $\beta$ to take the form
\begin{equation} \label{S}
ds^2=-\beta d \beta^2+R^{2}(\beta)(dx^2+dy^2+dz^2).
\end{equation}
It is now clear that the sign of $\beta$ determines the geometry,
being Lorentzian if $\beta>0$ and Euclidean if $\beta<0$. For
$\beta>0$, the traditional cosmic time can be recovered by the
substitution $t=\frac{2}{3}\beta^{3/2}$. Adapting the chart
$\{t,x,y,z\}$, we shall write $R(t)=R(\beta(t))$ and also
$\psi(t)=\psi(\beta(t))$. Using equations (\ref{N}) through
(\ref{S}) with units in which $\kappa=\hbar=c=1$ one finds the
following differential equations
\begin{equation} \label{T}
3\left(\frac{\dot{R}}{R}\right)^2-\Lambda=-V(\bar{\psi},\psi),
\end{equation}
\begin{equation} \label{U}
2\left(\frac{\dot{R}}{R}\right)^.+3\left(\frac{\dot{R}}{R}\right)^2-\Lambda=
\frac{1}{2}\left(\bar{\psi}\frac{\partial V}{\partial \bar{
\psi}}+\frac{\partial V}{\partial \psi}\psi
\right)-V(\bar{\psi},\psi),
\end{equation}
\begin{equation} \label{R}
\dot{\psi}+\frac{3}{2}\frac{\dot{R}}{R}\psi+\gamma^{0}\frac{\partial
V}{\partial \bar{\psi}}=0,
\end{equation}
\begin{equation} \label{V}
\dot{\bar{\psi}}+\frac{3}{2}\frac{\dot{R}}{R}\bar{\psi}-\frac{\partial
V}{\partial \psi}\gamma^0=0,
\end{equation}
where a dot represents differentiation with respect to $t$. We
formulate our differential equations in a region that does not
include $\beta=0$ and seek solutions for $R$ and $\psi$ that
smoothly pass through the $\beta=0$ hypersurface. The scalar
curvature corresponding to our metric is
\begin{equation} \label{Z}
{\cal
R}=6\left[\frac{\ddot{R}}{R}+\left(\frac{\dot{R}}{R}\right)^{2}\right],
\end{equation}
while the energy density of the spinor field is given by
\begin{equation} \label{X}
\rho=T_{00}=-V(\bar{\psi},\psi).
\end{equation}
It should be noted that equations (\ref{T})-(\ref{V}) are not all
independent. We shall see that with a suitable choice for the
potential $V(\bar{\psi},\psi)$, equation (\ref{U}) can be obtained
by combining equations (\ref{T}), (\ref{R}) and (\ref{V}). Also
equation (\ref{V}) is the Dirac equation for the adjoint spinor
$\bar{\psi}$ and should not be considered as an independent
equation. Thus, it is sufficient to solve either equations
(\ref{T}, \ref{R}) or equations (\ref{U} , \ref{R}) which we shall
endeavor to do in the next section after choosing a suitable form
for the potential $V(\bar{\psi},\psi)$.
\section{Solutions}
As mentioned in the last section integrability of the
Einstein-Dirac field equations (\ref{T})-(\ref{V}) depends on the
choice of a suitable form for $V(\bar{\psi},\psi)$. However, this
potential should also describe a physical self-interacting spinor
field. The potential is usually an invariant function constructed
from the spinor $\psi$ and its adjoint $\bar{\psi}$. In general,
difficulties in treating  Einstein-Dirac systems depend directly
on the form of the spinor $\psi$ and the potential
$V(\bar{\psi},\psi)$. For simplifying and addressing these
difficulties we first demand that $\psi$ should be a function of
$t$ only. This simplifies the Dirac equation (\ref{J}) into the
more manageable form  (\ref{N}). We should now  choose a suitable
self-interacting form for the potential. Some of the common forms
for $V$ are: $V(\bar{\psi},\psi)=m\bar{\psi}\psi$ representing a
free spinor field of mass $m$,
$V(\bar{\psi},\psi)=m\bar{\psi}\psi+J_{\mu}J^{\mu}$ where
$J^{\mu}=\bar{\psi}\gamma^{\mu}\psi$ and known as the Thirring
model, $V(\bar{\psi},\psi)=m\bar{\psi}\psi+(\bar{\psi}\psi)^2$
called Gross-Neveu model and
$V(\bar{\psi},\psi)=m\bar{\psi}\psi+(\bar{\psi}\psi)^2 -
(\bar{\psi}\gamma^{5}\psi)^2$ also known as the chiral Gross-Neveu
model \cite{10}. Since in this work we are going to study
signature transition for the solutions of the problem at hand, we
concentrate on the two of the simplest forms for
$V(\bar{\psi},\psi)$ which either give the exact solutions of the
field equations or show signature transition effects. We therefore
focus attention on the free spinor field and self-interacting
spinor field potentials with the latter having a term of the form
$(\bar{\psi}\psi)^2$.
\subsection{Free spinor field}
This is the case in which the potential has the form
$V(\bar{\psi},\psi)=m\bar{\psi}\psi$. From equation (\ref{P}) the
components of the energy-momentum tensor are
$T^0_0=m\bar{\psi}\psi$ and $T^i_i=0$, and our field equations
become
\begin{equation} \label{Y}
3\left(\frac{\dot{R}}{R}\right)^{2}-\Lambda=-m\bar{\psi}\psi,
\end{equation}
\begin{equation} \label{a}
2\left(\frac{\dot{R}}{R}\right)^{.}+3\left(\frac{\dot{R}}{R}\right)^{2}-\Lambda=0,
\end{equation}
\begin{equation} \label{b}
\dot{\psi}+\frac{3}{2}\frac{\dot{R}}{R}\psi+m\gamma^{0}\psi=0,
\end{equation}
\begin{equation} \label{c}
\dot{\bar{\psi}}+\frac{3}{2}\frac{\dot{R}}{R}\bar{\psi}-m\bar{\psi}\gamma^{0}=0.
\end{equation}
The solutions of this system can easily be obtained. The results
in terms of $\beta$ are listed as follows: \vspace{5mm}\noindent\\
{\bf Case (a)} {\bf $\Lambda=0$}: in this case integrating
equations (\ref{Y})-(\ref{c}) results in the solutions
\begin{equation} \label{d}
R(\beta)=\left(M\beta^{3/2}+A\right)^{2/3},
\end{equation}
\begin{equation} \label{e}
\psi=\left(%
\begin{array}{c}
  \psi_{+} \\
  \psi_{-} \\
\end{array}%
\right),\hspace{.5cm}
\psi_{\pm}\sim\frac{e^{\pm{2/3}im\beta^{3/2}}}{M\beta^{3/2}+A}.
\end{equation}
Also the scalar curvature ${\cal R}$ and energy density $\rho$ of
the spinor field are found to be
\begin{equation} \label{f}
{\cal R}=\frac{3M^2}{\left(M\beta^{3/2}+A\right)^2},\hspace{.5cm}
\rho=\frac{3M^2}{\left(M\beta^{3/2}+A\right)^2},
\end{equation}
where $M$ (being proportional to $m$) and $A$ are integrating
constants. We see that if $A\neq 0$, the scale factor $R(\beta)$
is not a real function in the Euclidean region $\beta<0$. If $A=0$
then $R(\beta)\sim \beta$ is an unbounded function in both
Euclidean and Lorentzian regions, passing continuously through
$\beta=0$. However, in this case the functions $\psi_{\pm}$,
${\cal R}$ and $\rho$ all have a singular behavior in $\beta=0$
but are well defined in the domains $\beta>0$ and $\beta<0$. The
solutions in the case of zero cosmological constant are therefore
not suitable candidates for exhibiting signature transition
behavior.\vspace{5mm}\noindent\\ {\bf Case (b)} {\bf $\Lambda<0$}:
in this case the solutions can be written as
\begin{eqnarray} \label{g}
R(\beta)=\left(\frac{M}{-\Lambda}\right)^{1/3}\cos^{2/3}
\left(\frac{\sqrt{-3\Lambda}}{3}\beta^{3/2}\right),
\end{eqnarray}
\begin{eqnarray} \label{h}
\psi_{\pm}\sim\frac{e^{\pm2/3im\beta^{3/2}}}{\sqrt{\frac{M}{-\Lambda}}
\cos\left(\frac{\sqrt{-3\Lambda}}{3}\beta^{3/2}\right)},
\end{eqnarray}
\begin{eqnarray} \label{i}
{\cal R}=-\Lambda
\left[\tan^{2}\left(\frac{\sqrt{-3\Lambda}}{3}\beta^{3/2}\right)-3\right],\hspace{.5cm}
\rho=\frac{-\Lambda}{\cos^{2}\left(\frac{\sqrt{-3\Lambda}}{3}\beta^{3/2}\right)},
\end{eqnarray}
where $M$ is an integrating constant proportional to  $m$, the
mass of the spinor field. The other integrating constant is chosen
so that $\dot{R}(\beta=0)=0$. The scale factor $R(\beta)$ in this
case is a real function and behaves exponentially for $\beta<0$,
passing smoothly through $\beta=0$ and becomes a bounded
oscillatory function for $\beta>0$. Also, the physical quantities
${\cal R}$, $\psi_{\pm}$ and $\rho$ are regular functions both in
the Euclidean and Lorentzian domains and pass continuously through
the $\beta=0$ hypersurface. Thus all of the above results show
that the solutions in the case of a negative cosmological constant
exhibit signature transition from a Euclidean to a Lorentzian
domain. A plot of the scale factor, scalar curvature and energy
density in terms of $\beta$ are shown in figure 1 through 3, using
equations
(\ref{g}) and (\ref{i}). \vspace{5mm}\noindent\\
{\bf Case (c)} {\bf $\Lambda>0$}: in this case for integrating the
system of equations (\ref{Y})-(\ref{c}) we must compare the Hubble
parameter $H=\frac{\dot{R}}{R}$ with the value of
$\sqrt{\frac{\Lambda}{3}}$. If $H^2>\frac{\Lambda}{3}$, the
solutions of the field equations in terms of the parameter $\beta$
read
\begin{equation} \label{j}
R(\beta)=\left(\frac{M}{\Lambda}\right)^{1/3}
\sinh^{2/3}\left(\frac{\sqrt{3\Lambda}}{3}\beta^{3/2}+A\right),
\end{equation}
where $M$, again proportional to the mass of spinor field $m$, and
$A$ are integrating constants. If we take $A=0$, the scale factor
$R(\beta)$ shows regular behavior in $\beta<0$ and $\beta>0$
regions and also pass smoothly through $\beta=0$. However, the
scalar curvature, energy density and spinor field
\begin{figure}
\centerline{\begin{tabular}{ccc}
\epsfig{figure=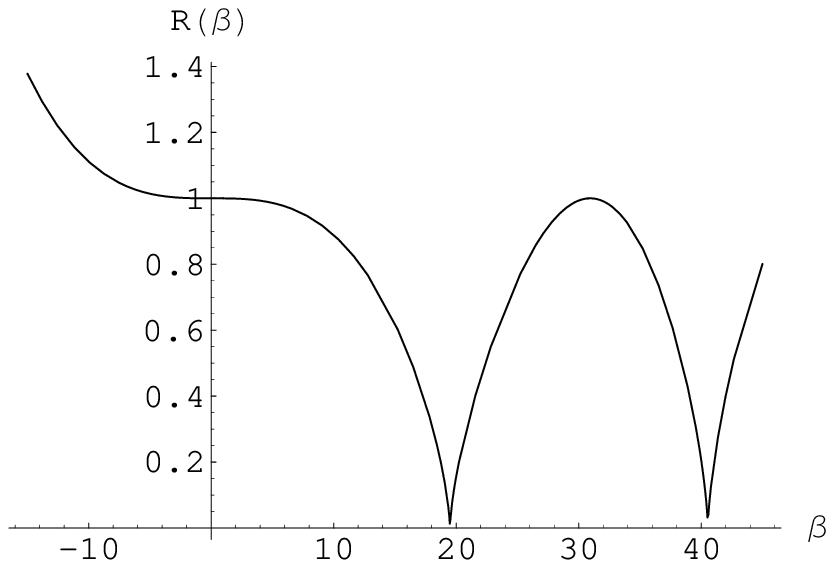,width=8cm}
 &\hspace{2.cm}&
\epsfig{figure=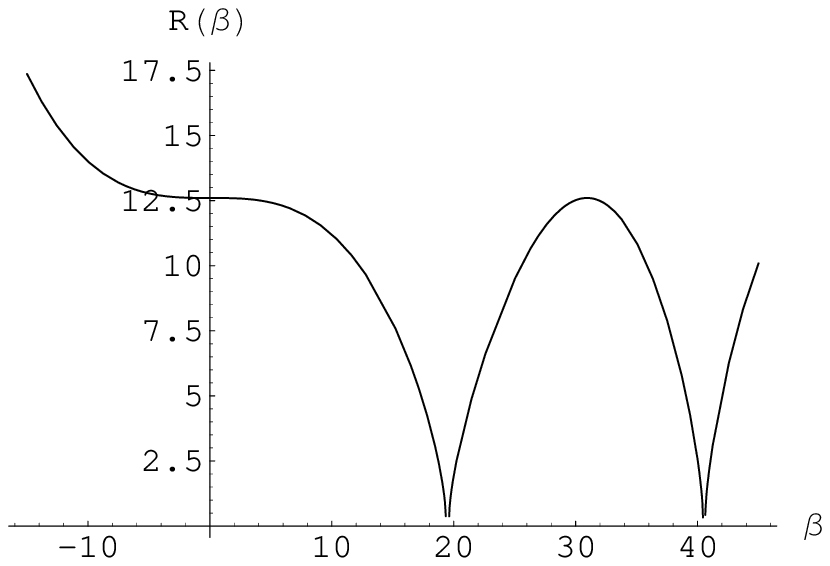,width=8cm}
\end{tabular}  }
\caption{\footnotesize Left, the scale factor for a free spinor
field  and right, the same figure  for a self interacting spinor
field, both for a typical negative value of $\Lambda=-10^{-3}$
with $M=g=1$.} \label{fig1}
\end{figure}
\begin{figure}
\centerline{\begin{tabular}{ccc}
\epsfig{figure=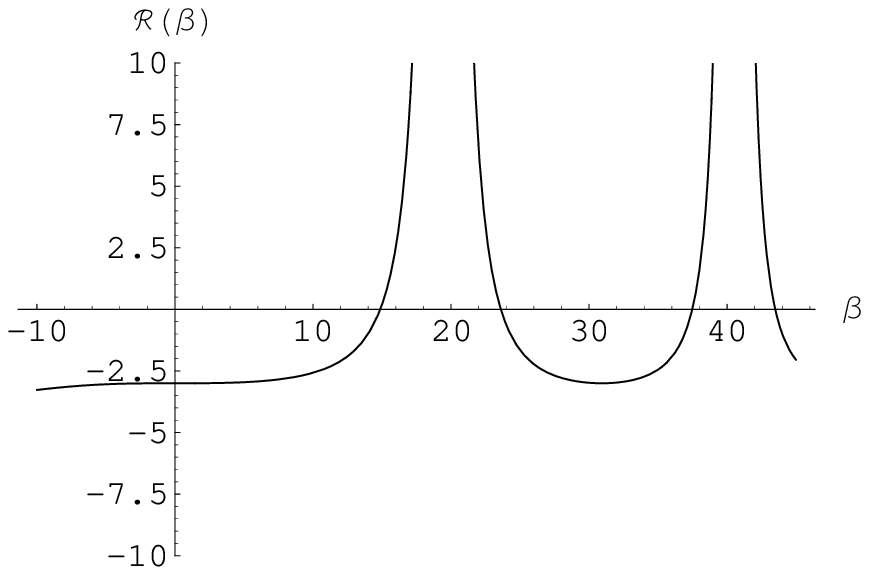,width=8cm}
 &\hspace{2.cm}&
\epsfig{figure=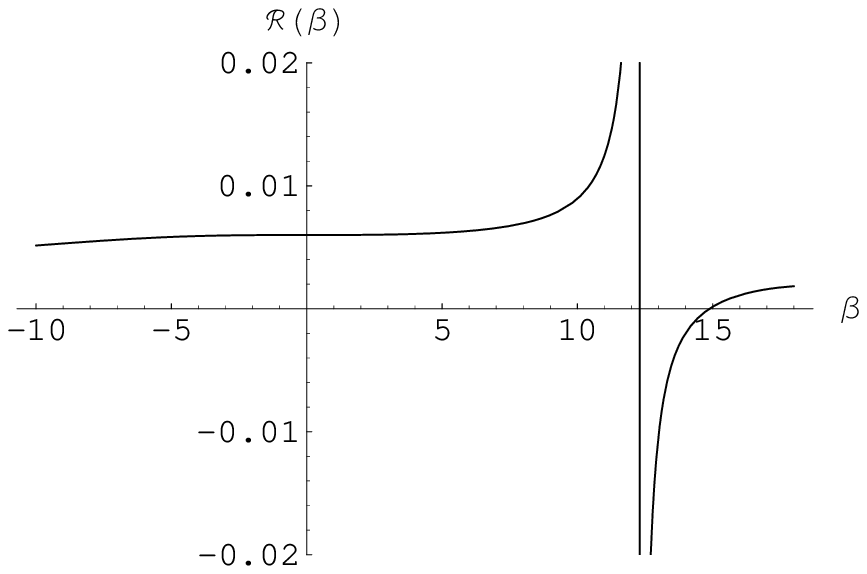,width=8cm}
\end{tabular}  }
\caption{\footnotesize Left, the curvature scalar for a free
spinor field and right, the same figure  for a self interacting
spinor field, both for a typical negative value of
$\Lambda=-10^{-3}$ with $M=g=1$.} \label{fig2}
\end{figure}
\begin{figure}
\centerline{\begin{tabular}{ccc}
\epsfig{figure=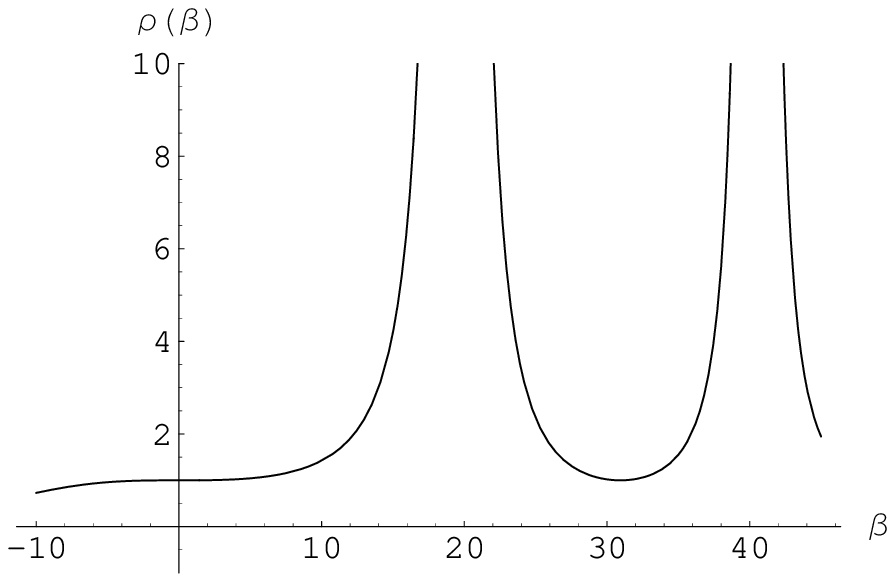,width=8cm}
 &\hspace{2.cm}&
\epsfig{figure=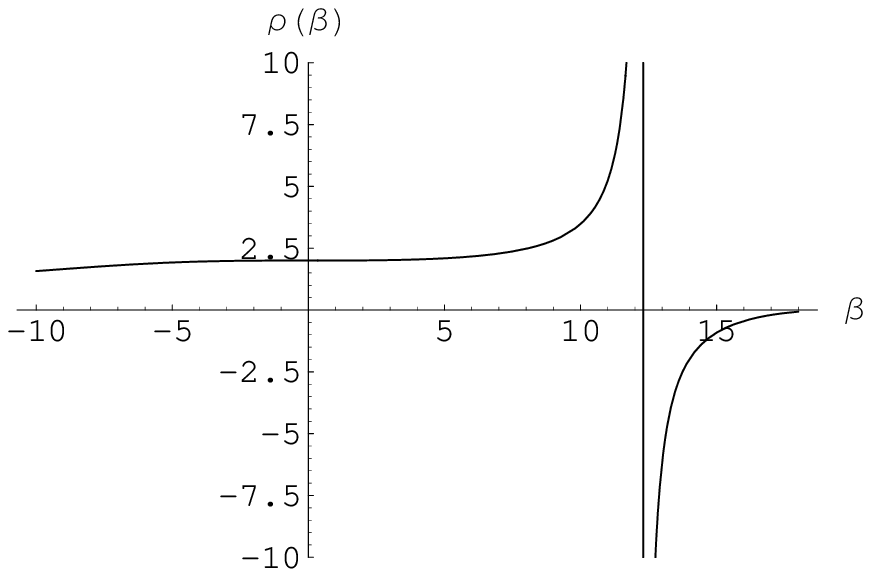,width=8cm}
\end{tabular}  }
\caption{\footnotesize Left, the energy density for a free spinor
field and right, the same figure  for a self interacting spinor
field, both for a typical negative value of $\Lambda=-10^{-3}$
with $M=g=1$.} \label{fig3}
\end{figure}

\begin{equation} \label{l}
{\cal R}=\Lambda
\left[\coth^{2}\left(\frac{\sqrt{3\Lambda}}{3}\beta^{3/2}\right)-3\right],
\end{equation}
\begin{equation} \label{m}
\rho=\frac{\Lambda}{\sinh^{2}\left(\frac{\sqrt{3\Lambda}}{3}\beta^{3/2}\right)},
\end{equation}
\begin{equation} \label{n}
\psi_{\pm}\sim\frac{e^{\pm2/3im\beta^{3/2}}}{\sinh
\left(\frac{\sqrt{3\Lambda}}{3}\beta^{3/2}\right)},
\end{equation}
all have singularity a at $\beta=0$. In the case $A\neq0$ the
scale factor (\ref{j}) is not a well defined real function when
passing through $\beta=0$. If $H^2<\frac{\Lambda}{3}$, the above
solutions take the form
\begin{equation} \label{o}
R(\beta)=\left(\frac{M}{\Lambda}\right)^{1/3}\cosh^{2/3}
\left(\frac{\sqrt{3\Lambda}}{3}\beta^{3/2}\right),
\end{equation}
\begin{equation} \label{p}
{\cal R}=\Lambda
\left[\tanh^{2}\left(\frac{\sqrt{3\Lambda}}{3}\beta^{3/2}\right)+3\right],
\end{equation}
\begin{equation} \label{r}
\rho=\frac{-\Lambda}{\cosh^{2}\left(\frac{\sqrt{3\Lambda}}{3}\beta^{3/2}\right)},
\end{equation}
\begin{equation} \label{s} \psi_{\pm}\sim\frac{e^{\pm2/3im\beta^{3/2}}}{
\cosh\left(\frac{\sqrt{3\Lambda}}{3}\beta^{3/2}\right)},
\end{equation}
where we have taken the integrating constant $A=0$. These
solutions resemble the ones for the  case $\Lambda<0$ and exhibit
signature transition. However,  we should note that the energy
density of the spinor field is negative and thus unphysical. In
summery, the above discussion shows that within the context of
this model, a universe with positive cosmological constant cannot
undergo signature transition from a Euclidean to a Lorentzian
domain through the $\beta=0$ hypersurface where its matter source
is a massive free spinor field.
\subsection{Self-interacting spinor field}
In this section we return to the field equations
(\ref{T})-(\ref{V}) with a potential describing a spinor field
interacting with itself. The form  which we have chosen for this
potential is that of the Gross-Neveu model
\begin{equation} \label{t}
V(\bar{\psi},\psi)=m\bar{\psi}\psi+\lambda(\bar{\psi}\psi)^{2},
\end{equation}
where $\lambda$ is a coupling constant. From (\ref{P}) the
components of the energy-momentum tensor read
\begin{equation} \label{u}
T^0_0=m\bar{\psi}\psi+\lambda(\bar{\psi}\psi)^2,\hspace{.5cm}
T^1_1=T^2_2=T^3_3=-\lambda(\bar{\psi}\psi)^2.
\end{equation}
The field equations (\ref{T})-(\ref{V}) can now be written as
\begin{equation} \label{x}
3\left(\frac{\dot{R}}{R}\right)^2-\Lambda=-m\bar{\psi}\psi-\lambda(\bar{\psi}\psi)^2,
\end{equation}
\begin{equation} \label{y}
2\left(\frac{\dot{R}}{R}\right)^{.}+3\left(\frac{\dot{R}}{R}\right)^2-\Lambda=
\lambda(\bar{\psi}\psi)^2,
\end{equation}
\begin{equation} \label{z}
\dot{\psi}+\frac{3}{2}\frac{\dot{R}}{R}\psi+(m+2\lambda
\bar{\psi}\psi)\gamma^{0}\psi=0,
\end{equation}
\begin{equation} \label{w}
\dot{\bar{\psi}}+\frac{3}{2}\frac{\dot{R}}{R}\bar{\psi}-(m+2\lambda
\bar{\psi}\psi)\bar{\psi}\gamma^0=0.
\end{equation}
If $\Lambda=0$, the scale factor as a function of $\beta$ obtained
from the above equations reads
\begin{equation} \label{AB}
R(\beta)=(M\beta^3-g)^{1/3},
\end{equation}
where $M$ and $g$ are two constants proportional to the mass of
the spinor field and the coupling constant $\lambda$ respectively.
The other integrating constant is taken to be zero. This function
shows an unbounded scale factor both in the Euclidean and
Lorentzian domains which passes continuously through $\beta=0$.
The scalar curvature and the energy density of the spinor field
are found to be
\begin{equation} \label{AC}
{\cal R}=-\frac{1}{3}M\frac{M\beta^{3}-3g}{(M\beta^{3}-g)^{2}},
\hspace{.5cm} \rho=\frac{3M^{2}\beta^{3}}{(M\beta^{3}-g)^{2}}.
\end{equation}
Functionally speaking, these are well defined in both $\beta<0$
and $\beta>0$ regions, passing smoothly through the signature
changing hypersurface. However, the question arises as to the
validity of these solutions since the energy density is again
negative in the Euclidean region ($\beta<0$), pointing to an
unphysical matter source. Like the free spinor field case, when
$\Lambda>0$, we are not led to physically well defined solutions.
In the case of $\Lambda<0$ the scale factor, being obtained from
the system of equations (\ref{x})-(\ref{w}) can be written as
\begin{equation} \label{AD}
R(\beta)=\left[\frac{M}{-\Lambda}+\sqrt{\frac{M^2}{\Lambda^2}+\frac{g}
{-\Lambda}}\cos\left(\frac{2}{3}\sqrt{-3\Lambda}\beta^{3/2}\right)\right]^{1/3},
\end{equation}
where as before $M$ and $g$ are equal to $m$ and $\lambda$ up to
an integrating constant, and the other integrating constant is
choose so that $\dot{R}(\beta=0)=0$. Also the Ricci scalar and the
spinor field energy density are obtained as follows
\begin{eqnarray} \label{AE}
{\cal
R}&=&\frac{6}{\frac{M}{-\Lambda}+\sqrt{\frac{M^2}{\Lambda^2}+
\frac{g}{-\Lambda}}\cos\left(\frac{2}{3}\sqrt{-3\Lambda}\beta^{3/2}\right)}\nonumber\\
&\times&\left[
\sqrt{M^{2}-g\Lambda}\cos\left(\frac{2}{3}\sqrt{-3\Lambda}\beta^{3/2}\right)+
\frac{\left(\frac{M^2}{-3\Lambda}+\frac{g}{3}\right)\sin^{2}\left(\frac{2}{3}
\sqrt{-3\Lambda}\beta^{3/2}\right)}{\frac{M}{-\Lambda}+\sqrt{\frac{M^2}
{\Lambda^2}+\frac{g}{-\Lambda}}\cos\left(\frac{2}{3}\sqrt{-3\Lambda}\beta^{3/2}\right)}\right]
\end{eqnarray}
\begin{equation} \label{AF}
\rho=\frac{1}{\frac{M}{-\Lambda}+\sqrt{\frac{M^2}{\Lambda^2}+
\frac{g}{-\Lambda}}\cos\left(\frac{2}{3}\sqrt{-3\Lambda}\beta^{3/2}\right)}
\left[2M+\frac{g}{\frac{M}{-\Lambda}+\sqrt{\frac{M^2}{\Lambda^2}+
\frac{g}{-\Lambda}}\cos\left(\frac{2}{3}\sqrt{-3\Lambda}\beta^{3/2}\right)}\right].
\end{equation}
A quick look at the functions (\ref{AD})-(\ref{AF}) shows their
regular behavior both in the Euclidean and Lorentzian regions
without any singularity at $\beta=0$. From (\ref{AD}) it is clear
that the corresponding cosmology is an unbounded universe in the
Euclidean domain having a bounded oscillatory behavior in the
Lorentzian region and passes smoothly through the hypersurface of
signature transition at $\beta=0$. Figure 1 through 3 show plots
of the scale factor, scalar curvature and energy density using
equations (\ref{AD}), (\ref{AE}) and (\ref{AF}).
\section{Massless spinor field}
At this point it would be interesting to take the massless limit
of the pervious results. If $m=0$ and the spinor field is free,
\textit{i.e.} $V(\bar{\psi},\psi)=m\bar{\psi}\psi$, then from
equations (\ref{Y})-(\ref{c}) we obtain
$R(\beta)=e^{\frac{2}{3}c\beta^{3/2}}$, where $c$ is a constant.
This is not a real function for $\beta<0$  and thus does not
exhibit signature transition. In the case of a self-interacting
spinor field however, the massless limit can be obtained easily by
substituting $M=0$ in equations (\ref{AD})-(\ref{AF}), which again
results in signature changing solutions. To summarize, solutions
exhibiting signature transition do not exist when the spinor field
is free and massless.
\section{Quantum cosmology}
The study of quantum cosmology of the model presented above is the
goal we shall pursue in this section. For this purpose we
construct the Hamiltonian of our model. Let us start with the
actin
\begin{equation}\label{AAA}
{\cal S}=\int \sqrt{-g}(L_{grav}+L_{matt})d^4 x,
\end{equation}
where
\begin{equation}\label{AAB}
L_{grav}={\cal R}-\Lambda,
\end{equation}
is the Einstein-Hilbert Lagrangian for the gravitational field and
$L_{matt}$ represents the Lagrangian of the matter source which is
given by (\ref{B}) with $g$ being the determinant of the metric
here. Now, by substituting (\ref{B}) and (\ref{AAB}) in
(\ref{AAA}) and integrating over spatial dimensions, we are led to
an effective Lagrangian in the mini-superspace
$\{R,\psi,\bar{\psi}\}$ as follows
\begin{equation}\label{AAC}
{\cal L}=R\dot{R}^2+\frac{1}{3}\Lambda
R^3+\frac{1}{6}R^3\left[\bar{\psi}\gamma^0\dot{\psi}-\dot{\bar{\psi}}
\gamma^0\psi-2V(\bar{\psi},\psi)\right].
\end{equation}
Variation of the above Lagrangian with respect to $R$,
$\bar{\psi}$ and $\psi$  yields  equations (\ref{U}), (\ref{R})
and (\ref{V}) respectively. Also, we have the ``zero energy
condition" given by
\begin{equation}\label{AAD}
{\cal H}=\frac{\partial {\cal L}}{\partial
\dot{R}}\dot{R}+\frac{\partial {\cal L}}{\partial
\dot{\psi}}\dot{\psi}+\dot{\bar{\psi}}\frac{\partial {\cal
L}}{\partial \dot{\bar{\psi}}}-{\cal L}=0,
\end{equation}
which yields the constraint equation (\ref{T}). To simplify the
Lagrangian (\ref{AAC}), consider the change of variable
$u=R^{3/2}$. In terms of this new variable Lagrangian (\ref{AAC})
takes the form
\begin{equation}\label{AAE}
{\cal L}=\frac{4}{9}\dot{u}^2+\frac{1}{3}\Lambda u^2
+\frac{1}{6}u^2\left[\bar{\psi}\gamma^0
\dot{\psi}-\dot{\bar{\psi}}\gamma^0
\psi-2V(\bar{\psi},\psi)\right],
\end{equation}
with the corresponding Hamiltonian given by
\begin{equation}\label{AAF}
{\cal H}=\frac{4}{9}\dot{u}^2-\frac{1}{3}\Lambda u^2
+\frac{1}{3}u^2 V(\bar{\psi},\psi)=0.
\end{equation}
Choosing the potential $V(\bar{\psi},\psi)=m
\bar{\psi}\psi+\lambda(\bar{\psi}\psi)^2$ and noting that
equations (\ref{z}) and (\ref{w}) allow us to write
$\bar{\psi}\psi=\frac{{\cal C}}{R^3}=\frac{{\cal C}}{u^2}$ and
substitute for the terms involving $\dot{\psi}$ and
$\dot{\bar{\psi}}$, we construct the Wheeler-DeWitt equation from
Hamiltonian constraint (\ref{AAF}) with the result
\begin{equation}\label{AAG}
{\cal
H}\Psi(u)=\left[\frac{d^2}{du^2}-\omega^2u^2+M-\frac{g}{u^2}\right]\Psi(u)=0,
\end{equation}
where $\omega^2=-\frac{16}{27}\Lambda$, $M=\frac{16}{27}m$,
$g=\frac{16}{27}\lambda$ and $\Psi(u)$ is the wave function of the
universe. Note that we have taken ${\cal C}=-1$ to obtain a
positive energy density for the spinor field, since in the case of
a positive cosmological constant, our classical solutions
(\ref{j}-\ref{s}) do not exhibit a physical situation. In what
follows, we only consider the case where $\Lambda<0$.

In order to solve equation (\ref{AAG}) we first consider the case
$g=0$, which is relevant to the free spinor field. In this case
equation (\ref{AAG}) becomes to a Hermite equation and its
eigenfunctions can be written in terms of Hermite polynomials
$H_n(x)$ as
\begin{equation}\label{AAI}
\Phi_n(u)=\left(\frac{\omega}{\pi}\right)^{1/4}\frac{1}{\sqrt{2^{n}n!}}e^{-\omega
u^2/2}H_n(\sqrt{\omega }u),\hspace{.5cm}n=0,1,2,\cdots
\end{equation}
The ``zero energy condition" ${\cal H}=0$ then yields
\begin{equation}\label{AAJ}
m=\frac{27}{16}(2n+1)\omega.
\end{equation}
We  note that the energy density of the spinor field in this case
is $\rho=-m\bar{\psi}\psi=\frac{m}{R^3}$. Thus, in our units, $m$
is the total energy of the spinor field. Therefore, equation
(\ref{AAJ}) is the quantization condition of the total energy of
the spinor field. The general solution of the WD equation in this
case can be written as
\begin{equation}\label{AAM}
\Psi(u)=\sum^\prime_n c_n\Phi_n(u),
\end{equation}
where the prime on the sum indicates summing over all values of
$n$ satisfying condition (\ref{AAJ}).

For a self-interacting spinor field, that is for $g\neq 0$,
equation (\ref{AAG}) after a change of variable $v=\omega u^2$ and
transformation $\psi=v^{-1/4}\phi$ becomes
\begin{equation}\label{AAN}
\frac{d^2\phi}{dv^2}+\left(-\frac{1}{4}+\frac{\kappa}{v}+
\frac{1/4-\mu^2}{v^2}\right)\phi=0,
\end{equation}
where $\kappa=\frac{M}{4\omega}$ and
$\mu^2=\frac{1}{16}+\frac{g}{4}$. The above equation is the
Whittaker differential equation and its solution can be written in
terms of confluent Hypergeometric functions $M(a,b,x)$ and
$U(a,b,x)$ as
\begin{equation}\label{AAO}
\phi(v)=e^{-v/2}v^{\mu+1/2}\left[c_1M(\mu-\kappa+\frac{1}{2},2\mu+1;v)+
c_2U(\mu-\kappa+\frac{1}{2},2\mu+1;v)\right].
\end{equation}
Bearing in the mind that if $g=0$ the self-interacting solution
should be reduced to solution (\ref{AAI}), we take $c_1=0$ and
retain only the functions $U(a,b,x)$ \cite{11}. Therefore, the
wave function $\Psi(u)$ is equal to
\begin{equation}\label{AAP}
\Psi(u)=\left(\omega
u^2\right)^{\frac{1+\sqrt{1+4g}}{4}}e^{-\omega
u^2/2}U\left(\frac{1}{2}-\frac{M}{4\omega}+\frac{\sqrt{1+4g}}{4},
1+\frac{\sqrt{1+4g}}{2}; \omega u^2\right).
\end{equation}
A glance at  equation (\ref{AAG}) shows that the solutions have a
behavior of the form $\Psi\sim e^{-\omega u^2/2}$ when
$u\rightarrow\infty$ and $\Psi\sim (u^2)^{(1+\sqrt{1+4g})/4}$ in
the limit where $u\rightarrow 0$. Demanding the same limiting
behavior for solutions (\ref{AAP}), one can easily see that the
function $U(a,b,x)$ in (\ref{AAP}) should reduce to a polynomial
and this happens when $a=-n$ \cite{11}. This yields a quantization
condition for the total energy of the self interacting spinor
field as follows
\begin{eqnarray}
m=\frac{27}{8}\left(2n+1+\frac{\sqrt{1+4g}}{2}\right). \label{eq1}
\end{eqnarray}

\begin{figure}
\centerline{\begin{tabular}{c} \epsfig{figure=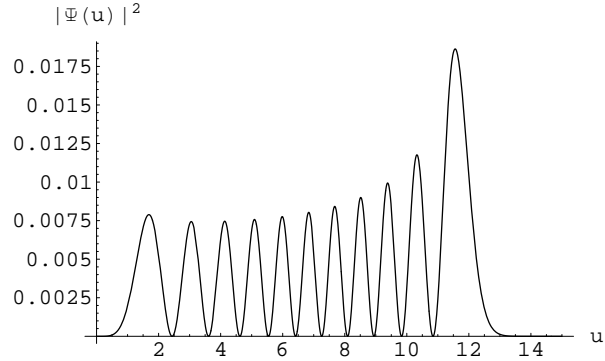,width=8cm}
\end{tabular}  }
\caption{\footnotesize Square of the wavefunction for $M=g=1$,
$\omega=0.025$ and $n=15$.} \label{fig4}
\end{figure}

Figure \ref{fig4} shows $|\Psi(u)|^2$ for typical values of the
parameters. As it is clear from this figure, the wavefunction has
a well defined behavior near $u=0$ and describes a universe
emerging out of nothing without any tunneling. Also the largest
peak in $|\Psi(u)|^2$ corresponds to the maximum of the scale
factor in the classical oscillatory solution as shown in figure
\ref{fig1}.
\section{Conclusions}
In this paper we have studied the properties of the Einstein-Dirac
equations in a spatially flat Robertson-Walker background, with an
eye to the signature changing solutions. We have shown that if the
cosmological constant is negative, either free or self-interacting
spinor fields result in solutions which admit a degenerate metric
in which the scale factor and the physical quantities of interest
such as scalar curvature and spinor field energy density have a
continuous behavior in passing from a Euclidean to a Lorentzian
domain. This phenomenon also appears when we have a
self-interacting massless spinor field. The corresponding
cosmology in these cases show an unbounded universe in the
Euclidean region passing smoothly through the signature changing
hypersurface and an oscillatory behavior in the Lorentzian domain.
If the cosmological constant is positive or zero the signature
changing solutions do not exist. In particular, equations
(\ref{o}-\ref{s}) show a universe with a positive cosmological
constant for which although signature transition occurs, the
energy density becomes negative which does not correspond to
ordinary matter. We have also shown that a massless free spinor
field does not result in such solutions, irrespective of the
choice of the potential.

The quantum cosmology of the model presented above and the ensuing
WD equation is amenable to exact solutions in terms of
Hypergeometric functions. These solutions correspond to the
classical signature changing solutions and would be of interest in
the context of the boundary condition of the universe.
\vspace{10mm}\noindent\\
{\bf Acknowledgement} The authors would like to thank the research
council of Shahid Beheshti university for financial support.

\end{document}